\documentclass[
reprint,
 amsmath,amssymb,
 aps,prl
]{revtex4-2}
\usepackage{amsmath}
\usepackage{braket}
\usepackage{float}
\usepackage{graphicx}
\usepackage{dcolumn}
\usepackage{bm}

\begin{document}

\preprint{APS/123-QED}

\title{Ultra-sensitive magnetic sensor based on 3-dimensional rotation induced Berry phase}
\author{Huaijin Zhang$^{1}$ and Zhang-Qi Yin$^{1}$}
\email{zqyin@bit.edu.cn}
\affiliation{$^{1}$Key Laboratory of Advanced Optoelectronic Quantum Architecture and Measurements (MOE), and Center for Quantum Technology Research, School of Physics, Beijing Institute of Technology}

\date{\today}

\begin{abstract}
High-sensitivity  magnetometers play a crucial role in various domains, including fundamental physics, biomedical imaging, and navigation. 
Levitated diamonds containing nitrogen-vacancy (NV) centers exhibit significant potential for magnetic sensing due to their high mechanical quality (Q) factor and long spin coherence time. 
However, previous studies have predominantly focused on electron spin-based measurements of alternating current (AC) magnetic fields. 
In this letter, we propose a novel approach for direct current (DC) magnetic field measurement based on the Berry phase generated by three-dimensional rotation. 
We analyze the adiabatic evolution of the $^{14}$N nuclear spin inside a  levitated 3D rotating diamond with frequencies around $\mathrm{MHz}$. 
Our finding reveals that the Berry phase exhibits high sensitivity to external parameters near rotation induced nuclear spin resonance. 
Using this mechanism, we theoretically demonstrate that the static magnetic field sensitivity can reach $10^{-7}~\mathrm{T/\sqrt{Hz}}/\sqrt{N}$ for $^{14}$N nuclear spins under the current experimental conditions.
\end{abstract}

\maketitle

\section{Introduction}
Precision measurement of magnetic fields serves as a cornerstone for progress in both fundamental \cite{crescini2022search} and applied physics \cite{lu2023recent}, enabling applications ranging from the investigation of condensed matter systems \cite{hossain2022synthesis} to biomedical imaging \cite{marmugi2019electromagnetic} and navigation technologies \cite{wang2022bioinspired}. 
In particular, static magnetic field sensing provides valuable information regarding material characteristics \cite{tetienne2014nanoscale} and biological phenomena \cite{le2013optical}. 
However, attaining both high sensitivity and micron-scale spatial resolution remains a significant technical challenge. 
Conventional sensing technologies, such as Hall probes and superconducting quantum interference devices (SQUIDs) \cite{vettoliere2023highly}, encounter inherent limitations in spatial resolution or operational requirements, thereby increasing the demand for quantum-based alternatives.

Quantum sensors based on atomic ensembles \cite{wasilewski2010quantum}, nitrogen-vacancy (NV) centers \cite{wolf2015subpicotesla, wang2022picotesla}, and cold atoms \cite{martin2017entanglement} have achieved remarkable sensitivity to dynamic magnetic fields, reaching levels on the order of $\mathrm{fT/\sqrt{Hz}}$. 
However, the measurement of static fields presents several unresolved trade-offs. 
Specifically, atomic vapor cells are limited by diffraction-limited spatial resolution ($>\mathrm{\mu m}$) \cite{sebbag2021demonstration}, which limits their applicability in smaller scale sensing. 
Although single NV electron spins allow nanoscale imaging \cite{grinolds2014subnanometre, guo2024wide}, their sensitivity to static fields is constrained by rapid dephasing \cite{chen2023extending, barry2020sensitivity}. 
Recent studies have used the nuclear spins of $^{14}$N or $^{13}C$ as quantum memories to improve the sensing performance \cite{voisin2024nuclear}. 
Despite these advances, current approaches to nuclear spin manipulation rely predominantly on microwave or radiofrequency excitation \cite{goldman2020optical, wood2021quantum}, which can lead to power dissipation and increased decoherence. 
Furthermore, geometric phase-based schemes for measuring static magnetic fields remain largely unexplored, mainly due to the low gyromagnetic ratios of nuclear spins ($\gamma_n\approx10^{-4}\gamma_e$) and challenges in achieving high-fidelity control.

Here, we propose a magnetometer based on the 3D rotation-induced Berry phase. 
By 3-dimensionally rotating a levitated diamond, the embedded NV center spin states undergo periodic evolution. 
After treating the influence of electron spin through perturbation theory, we obtain the Hamiltonian of the $^{14}$N nuclear spin in 3D rotation frame.
We analyze the range of rotational parameters necessary to achieve nuclear spin resonance conditions, and provide the relationship between the Berry phase and the rotation frequency. 
The accumulated Berry phase exhibits high sensitivity to external parameters near the nuclear spin resonance boundary, providing a noise-resistant measurements. 
In the absence of external magnetic driving fields, the rotational parameters, particularly the rotational frequency, can preserve system adiabaticity across a wide operational range. 
A single $^{14}$N spin demonstrates a sensitivity of approximately $10^{-7}~\mathrm{T/\sqrt{Hz}}$ under millitesla-level static magnetic fields, while offering micron-scale spatial resolution.

\section{Theory}
\begin{figure}[htbp]
    \centering
    \includegraphics[width=8.6cm]{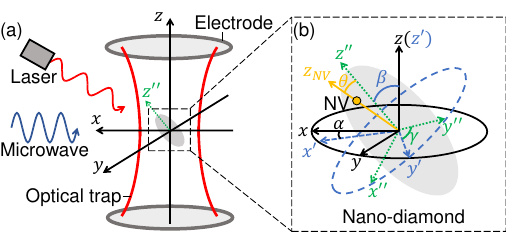}
    \caption{(a) A non-spherical diamond is levitated in a vacuum via a hybrid trapping system. The spin states of the NV center are manipulated using microwave fields, while the motion of the diamond is detected through laser. Precise control over both the translational and rotational degrees of freedom of the diamond can be achieved by fine-tuning optical traps and electrodes. (b) The laboratory frame, represented as ${x,y,z}$, and the body frame, designated as ${x'',y'',z''}$, can be mutually transformed using a set of Euler angles ${\alpha,\beta,\gamma}$. Furthermore, an angle $\theta$ exists between the NV axis and the $z''$ axis. }
    \label{fig:1}
\end{figure}

We propose a scheme that a diamond is levitated in a vacuum through a hybrid trap \cite{jin2024quantum}, as depicted in Fig. \ref{fig:1}(a).   
The translation and rotation of the diamond can be precisely controlled using both optical traps and electrodes. 
The frequency of diamond translational motion $\sim \mathrm{kHz}$ is substantially lower than that of rotation $\sim \mathrm{MHz}$.
Therefore the coupling between the rotational and translational motion can be neglected. 
Besides, both the relativistic effects and deformation of the diamond are also neglected.  
We utilize the $3$D rotational operator \( R = R_{z}(\alpha)R_{y}(\beta)R_{z}(\gamma)R_{y}(\theta) \) to describe the rotation of the diamond \cite{chen2019nonadiabatic, rashid2018precession}, where $R_{i}(\psi) = \exp [- \mathrm{i}\psi (I_{i}+S_{i})]$ denotes the rotation operator along the $i=x,y,z$ axis, and $\psi$ represents the angle of rotation. 
Here, $\{\alpha, \beta, \gamma\}$ are the Euler angles, which define the relative orientation between the laboratory frame $\{x,y,z\}$ and the body frame $\{x'',y'',z''\}$, as illustrated in Fig. \ref{fig:1}(b). The electron spin is represented by $S$, while the nuclear spin is denoted by $I$.

In the laboratory frame, 
the Hamiltonian of a single NV center can be expressed as \cite{jarmola2020robust}: $H=R(t)H_0R^\dagger(t)+\gamma_e B\cdot S+\gamma_n B\cdot I$, where $H_0=DS_z^2+QI_z^2+A_{\parallel}S_{z}I_{z}+(A_{\perp}/2)\left(S_{+}I_{-}+S_{-}I_{+}\right)$. 
Here, $D$ represents the zero-field splitting of the NV center electron spin states, $Q$ denotes the nuclear quadrupole coupling constant, while $A_\parallel$ and $A_\perp$ correspond to the axial and transverse magnetic hyperfine coupling constants, respectively. 
Additionally, $B$ refers to a static magnetic field, $\gamma_e$ and $\gamma_n$ are the gyromagnetic ratios of the electron and nucleus, respectively. 
For simplicity, we set $\hbar=1$. 

\begin{figure}[htbp]
    \centering
    \includegraphics[width=8.6cm]{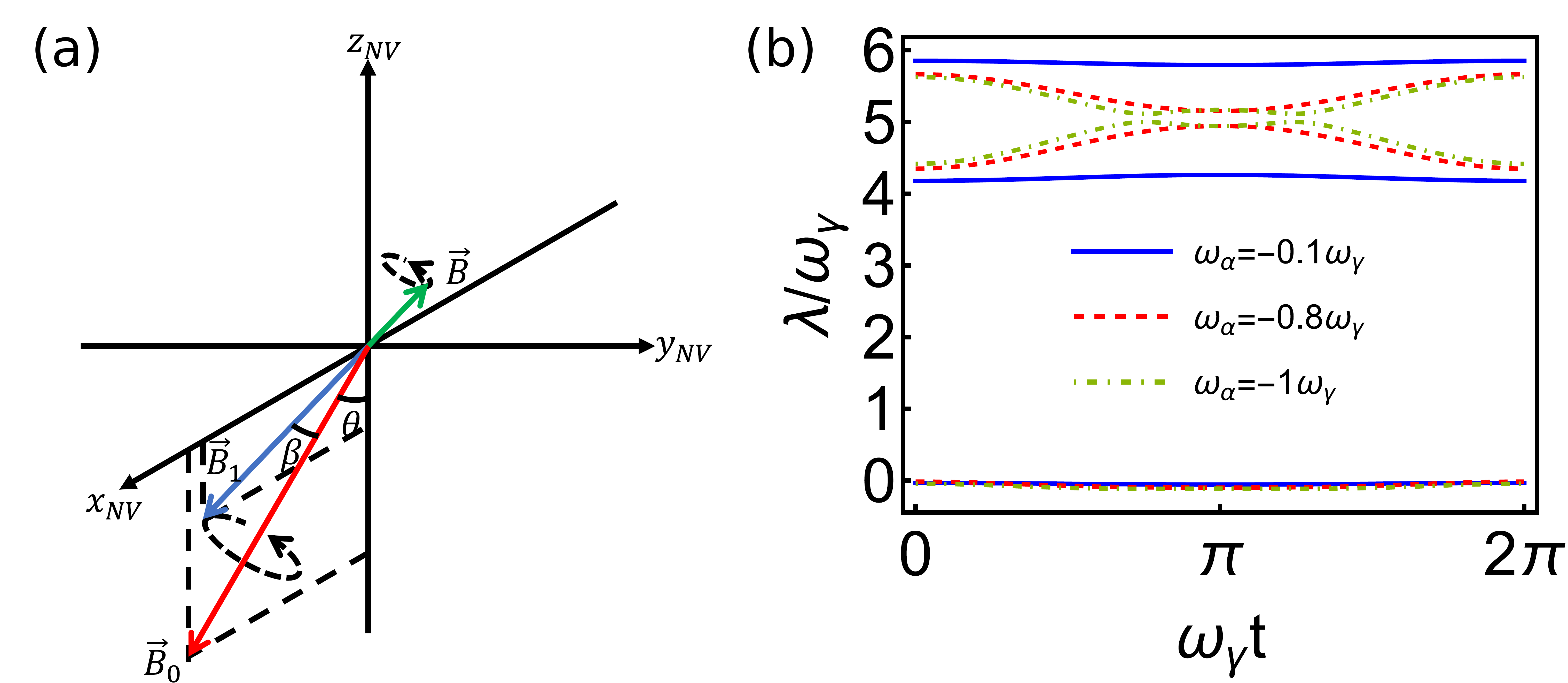}
    \caption{(a) In the body frame, $\vec{B}_0$ and $\vec{B}_1$ are the effective magnetic fields induced by rotation, and $\vec{B}$ is the static magnetic field in the inertial system. $\vec{B}_0$ is the static magnetic field in the $z-x$ plane with a magnitude of $\omega_\gamma/\gamma_n$ and an angle of $\pi-\theta$ with the $z$-axis. The magnitude of $\vec{B}_1$ is $\omega_\alpha/\gamma_n$, the angle between $\vec{B}_1$ and $\vec{B}_0$ is $\beta$, and $\vec{B}_1$ rotates around $\vec{B}_0$ with a frequency of $\omega_\gamma$. The $\vec{B}$ always in the opposite direction of $\vec{B}_1$. (b) The variations of the eigenvalues within one period for different $\omega_\alpha$. The curves of the same color from top to bottom are $\lambda_1$, $\lambda_2$, and $\lambda_3$ respectively. }
    \label{fig:2}
\end{figure}

We assume that the magnetic field $B$ is aligned along the $z$-axis. 
Additionally, considering that nutation can take negative values, the Euler angles are expressed as $\alpha = \omega_\alpha t$, $\beta = \beta_0$, and $\gamma = \omega_\gamma t$. 
In this letter, we investigate the magnetic field sensing utilizing the $^{14}$N nuclear spin, which requires $\omega_\alpha$ ($\omega_\gamma$) satisfying $\lesssim Q$ and $\gamma_nB \ll Q$.
Since $D \ll Q$, by fixing the NV center electron spin to $S_z = 0$, perturbation theory can be employed to decouple the electron spin from the nuclear spin. 
This results in a shift of $Q$, such that $Q' = Q + A_{\perp}^2 / D$. 
Subsequently, we utilize the operator $W(t) = R^{\dagger}(t)$ to derive the interaction Hamiltonian in the form $H_I = W H W^\dagger + \mathrm{i} \partial_t W W^\dagger$. 
The effective interaction Hamiltonian of the nuclear spin is:
\begin{align}
    \label{eq1}
    H_{In}=Q'I_z^2-\omega_\alpha' R^\dagger(t)I_zR(t)-\omega_\gamma R^\dagger_y(\theta) S_z R_y(\theta),
\end{align}
where $\omega_\alpha'=\omega_\alpha -\gamma _nB$.
The effect of rotation is equivalent to a static magnetic field $\vec{B}_0 = - (\omega_\gamma/\gamma_n) (-\sin{\theta} \vec{e}_{x_{NV}} + \cos{\theta} \vec{e}_{z_{NV}})$ and a rotating magnetic field $\vec{B}_1 = - (\omega_\alpha/\gamma_n) [-(\cos{\beta}\sin{\theta}+\sin{\beta}\cos{\theta}\cos{\omega_\gamma t}) \vec{e}_{x_{NV}}+(\sin{\beta}\sin{\omega_\gamma t}) \vec{e}_{y_{NV}}+(\cos{\beta}\cos{\theta}-\sin{\beta}\sin{\theta}\cos{\omega_\gamma t}) \vec{e}_{z_{NV}}]$. 
Furthermore, in the body frame of the NV center, the magnetic field $\vec{B}$ also rotates around $\vec{B}_0$ and is antiparallel to $\vec{B}_1$: $\vec{B}=-(\gamma_nB/\omega_\alpha)\vec{B}_1$. 
All these effective magnetic fields are depicted in Fig. \ref{fig:2}(a).

The matrix representation of the Hamiltonian \eqref{eq1} is given by: 
\begin{align}
    \label{eq2}
    H_{In}=\left[
    \begin{array}{ccc}
         Q'+\gamma_n B'_z&\frac{\gamma_n(B'_x-\mathrm{i}B'_y)}{\sqrt{2}}&0  \\
         \frac{\gamma_n(B'_x+\mathrm{i}B'_y)}{\sqrt{2}}&0&\frac{\gamma_n(B'_x-\mathrm{i}B'_y)}{\sqrt{2}} \\
         0&\frac{\gamma_n(B'_x+\mathrm{i}B'_y)}{\sqrt{2}}&Q'-\gamma_n B'_z
    \end{array}
    \right],
\end{align}
where $\vec{B}'=\vec{B}+\vec{B}_0+\vec{B}_1$. 
When the $z$-component of the magnetic field significantly exceeds the transverse component, transitions between different spin states are effectively suppressed. 
Conversely, when the transverse component dominates over the $z$-component, the spin states $\ket{I_z=+1}$ and $\ket{I_z=-1}$ become nearly degenerate, leading to Rabi resonance between these two states. 
Given that the vectors $\vec{B}_0$ and $\vec{B}_1$ can be controlled by adjusting the rotational parameters, it is always possible to identify a set of suitable parameters such that the total magnetic field exhibits only a transverse component at a specific moment.
Actuality, these parameters are only required to satisfy the following condition:
\begin{align}
    \label{eq3}
    \left[\frac{\omega'_\alpha}{\omega_\gamma}\cos{(\theta+\beta)}+\cos{\theta}\right]\left[\frac{\omega'_\alpha}{\omega_\gamma}\cos{(\theta-\beta)}+ \cos{\theta}\right]\leq0.
\end{align} 
We focus on the critical situation: $\omega_\alpha \cos{(\theta-\beta)}=-\omega_\gamma \cos{\theta}$.  
This condition corresponds to the occurrence of degeneracy at $t = T/2$, where $T=2\pi/\omega_\gamma$ denotes the period of the $H_{In}$.

We denote the eigenstates of the Hamiltonian $H_{In}$ as $\ket{\lambda_n}$, where $\lambda_n$ denotes the eigenvalue of $H_{In}$.  
By fixed $\beta$, $\theta$ and $\omega_\gamma$,  the variation of eigenvalues over time for different ratios of $\omega_\alpha/\omega_\gamma$ has been plotted, as illustrated in Fig. \ref{fig:2}(b).  
According to Eq. \eqref{eq3}, the critical value of $\omega_\alpha/\omega_\gamma$ satisfying the resonance condition is calculated to be $-\cos{\theta}/\cos{(\theta-\beta)}$. 
This indicates that as $|\omega_\alpha/\omega_\gamma|$ increases, the system transitions from states of large detuning to near-resonance, and to complete resonance. 
Nevertheless, over a given time of period, even for systems fulfilling the resonance condition, it remains a substantial discrepancy between the initial eigenvalues. 
As time progresses, two of these eigenvalues begin to approach each other and eventually cross at the resonance point. 
As shown in Fig. \ref{fig:2}(b), under non-critical conditions (such as $\omega_\alpha=-\omega_\gamma$) for resonance, this process occurs twice. 

\begin{figure}[htbp]
    \centering
    \includegraphics[width=8.6cm]{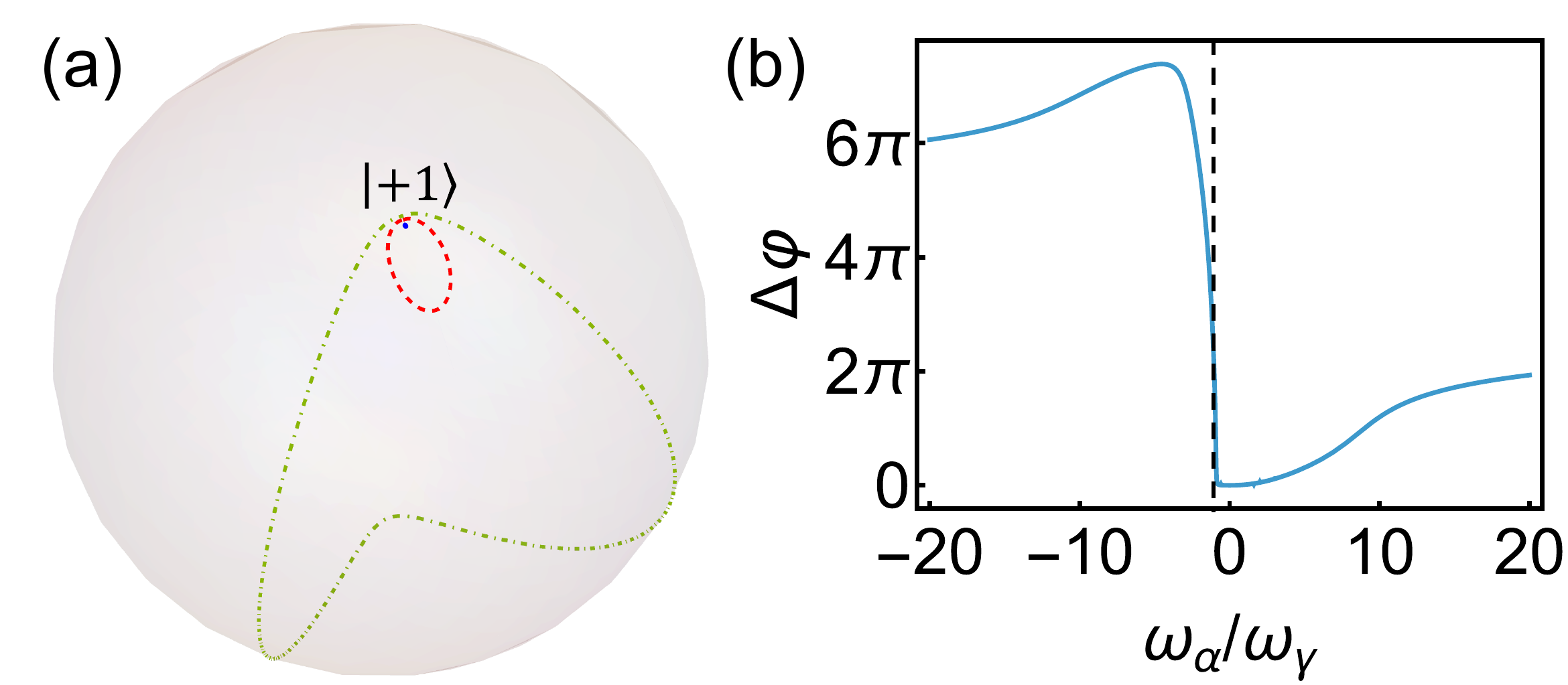}
    \caption{(a) The eigenstate $\ket{\lambda_1}$ corresponding to different $\omega_\alpha$ on the Bloch sphere. The two poles of the sphere are the spin states $\ket{+1}$ and $\ket{-1}$ respectively. (b) The geometric phase shift $\Delta \varphi_g$ corresponding to different $\omega_\alpha$. The maximum slope is obtained at the dotted line ($\omega_\alpha=-[\cos{\theta}/\cos{(\theta-\beta)}]\omega_\gamma$). }
    \label{fig:3}
\end{figure}

We assume that the system evolves adiabatically, the following condition is satisfied:
\begin{align}
    \label{eq4}
    \epsilon_{mn}=\left|\frac{\bra{\lambda_m}\dot{H}_{In}\ket{\lambda_n}}{(\lambda_n-\lambda_m)^2}\right|<<1. 
\end{align} 
The system will stay at the instantaneous eigenstate $\ket{\lambda_n}$, and accumulate a phase factor $\exp{(\mathrm{i}\varphi_n)}$. 
Here, the phase $\varphi_n$ consists of two components: the dynamic phase $\varphi_{dn}(t)=-\int_0^t\lambda_n(t')dt'$ and the Berry phase $\varphi_{gn}(t)=\mathrm{i}\int_0^t\bra{\lambda_n(t')}\partial_{t'}\ket{\lambda_n(t')}dt'$. 
Notably, in near-resonance systems, the adiabatic condition imposes strict constraints on the parameters, which will be elaborated in the subsequent section. 
Herein, we present the Berry phase accumulated over a single period $T$.
Firstly, we rewrite the eigenstates as $\ket{\lambda_n}=[\sin{(\vartheta_{1n}/2)} \sin{(\vartheta_{2n}/2)} \exp{(\mathrm{-i\phi})}]\ket{+1}+ \cos{(\vartheta_{1n}/2)\ket{0}+[\sin{(\vartheta_{1n}/2)} \cos{(\vartheta_{2n}/2)} \exp{(\mathrm{i\phi})}}]\ket{-1}$.
The range for both $\vartheta_{1n}$ and $\vartheta_{2n}$ is $[0,\pi]$. 
$\phi$ is the argument of $B_x'+\mathrm{i}B_y'$. 
Subsequently, utilizing these parameters, the Berry phase can be formulated as $\varphi_{ng}=(1/2)\int_0^T\dot{\phi}(1-\cos{\vartheta_{1n}})\cos{\vartheta_{2n}}\mathrm{d}t$.
Since we assume that the spin states $\ket{\pm1}$ are degenerate, it follows that $|\braket{0|\lambda_n}| \ll 1$ for $n=1, 2$. 
Consequently, we obtain $\vartheta_{1n} \approx \pi$, which implies that the eigenstate is approximately given by $\ket{\lambda_n} = \exp{(-i\phi)}\sin{(\vartheta_{2n}/2)}\ket{+1} + \exp{(i\phi)}\cos{(\vartheta_{2n}/2)}\ket{-1}$. 
The eigenstates can be visualized on the Bloch sphere, as depicted in Fig. \ref{fig:3}(a). 
The Berry phase corresponds to the solid angle subtended by the trajectory of the eigenstates. 
The dynamical phase shift $\varphi_{d1}-\varphi_{d2}$ approaches to $0$.

The system resonance condition is governed by the parameters $\omega_\gamma$, $\omega_\alpha'$, $\theta$, and $\beta$. 
By fixing $\omega_\gamma$, $\theta$, and $\beta$, the Berry phases become dependent solely on $\omega_\alpha'$. 
Near resonance, even a minor variation in this parameter can result in a significant change in the Berry phase, as illustrated in Fig. \ref{fig:3}(b). 
And this slope remains constant within the range of $\delta\omega_\alpha'/\omega_\alpha'<<1$, which means the geometric phase is linearly depend on $\omega_\alpha'$ ($B$).
Based on this mechanism, we propose a magnetic field measurement scheme.

\section{measurement sensitivity} 
Although the Berry phase of the system demonstrates maximal sensitivity to parameter variations near resonance, fulfilling the adiabatic condition requires precise control over the parameters.   
According to Fig. \ref{fig:3}(b), it is observed that when the system reaches resonance at $t = T/2$, the Berry phase exhibits its highest sensitivity.
At this stage, the primary constraint on the parameters predominantly arises from the condition $\epsilon_{12} \ll 1$.
The rotation parameters satisfy the relationship $\omega_\alpha = -\omega_\gamma \cos{\theta} / \cos{(\beta - \theta)}$. 
Thus, the adiabatic condition  \eqref{eq4} can be mathematically expressed as follows:
\begin{align}
    \label{eq5}
    \epsilon_{12}=\left|\frac{\frac{\sin{\beta}\cos{\theta}}{\cos{(\beta-\theta)}}}{\sqrt{2\sqrt{\chi^2+\zeta^2}[\sqrt{\chi^2+\zeta^2}-\chi]^3}}\right|<<1;
\end{align} 
where $\chi = |Q' / 2\omega_\gamma|$.

The Berry phase can be precisely measured by Ramsey interference \cite{soshenko2021nuclear}.
We assume that the nuclear spin of the NV center is initially prepared in the state $\ket{0}$.
A $\pi/2$ pulse transforms the nuclear spin state to the superposition state $(\ket{+1}+\ket{-1})/\sqrt{2}$ \cite{wood2021quantum, neumann2010single, dutt2007quantum}. 
Because the NV centers are within a region around $1~\mathrm{\mu m^2}$ during the rotation process, the pulse can continuously polarize the nuclear spins. 
The pulse duration is approximately $10~\mathrm{\mu s}$, which is much shorter than the measurement time of $T_m=10~\mathrm{ms}$. 
The final state of the system is given by $(\ket{+1}+\exp{(\mathrm{i}\Delta\Psi)}\ket{-1})/\sqrt{2}$. 
At this stage, an additional $\pi/2$ pulse is applied.
By measuring the fluorescence signal corresponding to the spin state $\ket{0}$ and observing the resulting interference fringes, the phase $\Delta\Psi$ can be accurately determined.
The measurement accuracy of the phase in this process is fundamentally limited by shot noise \cite{taylor2008high, budker2007optical}, resulting in an uncertainty of $\delta\Delta\Psi\sim1/\sqrt{N}$.
Here, $N$ denotes the total number of NV center nuclear spins involved in the measurement, and these NV centers are required to share the same orientation.

\begin{figure}[htbp]
    \centering
    \includegraphics[width=8.6cm]{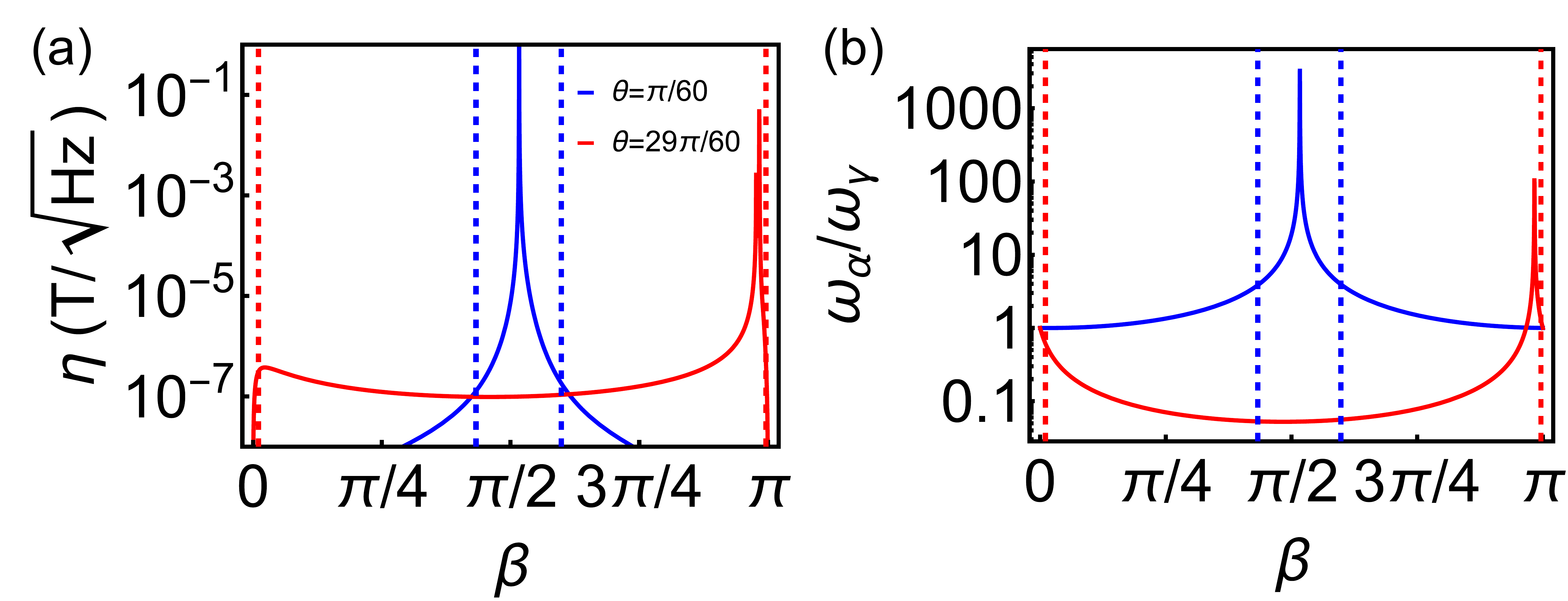}
    \caption{(a) The curves of sensitivity $\eta$ as a function of angle $\beta$ at various angles $\theta$. Here, we assume that $\chi=1$ and $\omega_\alpha=-\cos{\theta}/\cos{(\beta-\theta)}$. The dotted lines in different colors correspond to the boundaries of the adiabatic conditions for different values of $\theta$. Only the regions between these dotted lines satisfy the adiabatic condition. (b) The curves of $\omega_\alpha$ as a function of $\beta$, plotted under the same parameters.}
    \label{fig:4}
\end{figure}
 
The total geometric phase shift $\Delta\Psi$ is related as $\delta\Delta\Psi=(T_m/T)(\partial\Delta\varphi_g/\partial\omega_\alpha')\gamma_n\delta B$.
Thus, the measurement sensitivity can be expressed as follows:
\begin{align}
    \label{eq6}
    \eta=\delta B\sqrt{T_m}=\frac{2\pi}{\gamma_n \sqrt{NT_m}}\left[\frac{\partial\Delta\varphi_g}{\partial(\omega_\alpha'/\omega_\gamma)}\right]^{-1}
\end{align} 
The magnetic field measurement sensitivity of a single $^{14}$N nuclear spin is estimated to be on the order of $\sim10^{-7}~\mathrm{T/\sqrt{Hz}}$, as illustrated in Fig. \ref{fig:4} (a). 
We specifically chose two representative angles, $\theta \approx 0$ and $\theta \approx \pi/2$, to elucidate the relationship among sensitivity, adiabatic conditions, and $\beta$. 
Firstly, singular points emerge in the $\eta-\beta$ plane and shift to the right with an increase in the angle $\theta$. 
As shown in Fig. \ref{fig:4}, these singular points correspond to the scenario where $\beta-\theta$ approaches $\pi/2$, causing $\omega_\alpha$ to tend toward infinity. 
For comparison, we also plotted the variation of $\omega_\alpha$ with respect to $\beta$ under the same parameters in Fig. \ref{fig:4} (b).
From the figure, it is evident that the singular point of sensitivity coincides with the divergence point of the rotational frequency $\omega_\alpha$.
In addition, under adiabatic conditions, the boundaries are indicated by dotted lines in the corresponding color.
Only the parameters located between the two dashed lines of the same color satisfy the adiabatic condition.
Therefore, we observe that as $\theta$ increases, the distance between the dashed lines also increases. When $\theta$ approaches $\pi/2$, nearly all values of $\beta$ can satisfy the adiabatic condition.
Finally, it can be concluded that when $\theta \approx \pi/2$, the optimal sensitivity is achieved at $\beta \approx \pi/2$. At this point, the measurement sensitivity for a single nuclear spin can reach $\eta = 10^{-7}~\mathrm{T/\sqrt{Hz}}$.
Meanwhile, the rotation frequencies are given by $\omega_\gamma \approx Q'/2 = 2\pi \times 2.5~\mathrm{MHz}$ and $\omega_\alpha \approx 0.05\omega_\gamma = 2\pi \times 125~\mathrm{kHz}$. 
The measured magnetic field needs to meet the condition $|B|<<\omega_\alpha/\gamma_n\sim10~\mathrm{mT}$, this indicates that our measurement scheme is applicable to the measurement of static magnetic fields in the $\sim\mathrm{mT}$ range.

\section{Feasibility and Noise analyze} 
As previously discussed, the dominant source of noise in the measurement process is shot noise. 
To mitigate this, an ensemble of nuclear spins can be utilized for enhanced measurement sensitivity. 
It is essential that the principal axes of these NV centers are aligned at the same angle relative to the principal axis of the diamond crystal \cite{suto2017highly}. 
Under such conditions, the measurement sensitivity scales with the square root of the number $N$ of participating spins, improving to $1/\sqrt{N}$. 
Additionally, spin-state dephasing \cite{ajoy2012stable} can significantly degrade measurement accuracy. 
However, the dephasing time of the $^{14}$N nuclear spin can extend to approximately several seconds. 
Given that the magnetic field $B$ being measured constitutes an alternating current (AC) field in the body frame, techniques such as spin echo \cite{delord2018ramsey} can be applied to effectively suppress this noise \cite{wood2018t}. 
Therefore, the influence of dephasing on the system can be regarded as negligible.

Our scheme requires that the rotation frequency of the levitated diamond match the characteristic frequency of the nuclear spin. 
Recent experiments have demonstrated rotational frequencies on the order of $\sim \mathrm{GHz}$ for $100~\mathrm{nm}$-sized particles \cite{reimann2018ghz, jin20216}, and approximately $\sim10~\mathrm{MHz}$ for micrometer-sized particles \cite{kuhn2017optically, jin2024quantum}. 
Rotational instability also constitutes a significant source of noise. 
The heating caused by collisions with residual gas molecules, as well as intrinsic rotational instabilities, can both lead to fluctuations in rotational parameters: $\delta\omega_\alpha$, $\delta\omega_\gamma$, and $\delta\beta$. However, as illustrated in Fig. \ref{fig:4}(a), under the selected parameter settings, the measurement sensitivity exhibits minimal sensitivity to small perturbations in $\beta$. 
Thermal motion-induced noise can be mitigated through the application of feedback cooling techniques \cite{tebbenjohanns2021quantum, magrini2021real, delic2020cooling}. 
Furthermore, high rotational frequencies such as $\omega_\alpha$ and $\omega_\gamma$ can be effectively suppressed using feedback control mechanisms \cite{kuhn2017optically}, reducing their disturbances to a level of $\delta\omega/\omega \sim 10^{-12}$. 
The resulting impact of this suppressed rotational instability on magnetic field measurement sensitivity is estimated to be approximately $\delta\eta \sim 10^{-13}~\mathrm{T/\sqrt{Hz}}$, which is considered negligible.

\section{Conclusion}   
We investigate the dynamics of the $^{14}$N nuclear spin coupled to the NV center in a three-dimensional rotating diamond. 
In the body frame of the NV center, the 3D rotation is equivalent to applying a rotational magnetic field. 
We calculate the adiabatic geometric phases induced by the 3D rotation on the nuclear spin states. 
Our results demonstrate that this adiabatic geometric phase exhibits high sensitivity to the magnetic field near resonance. 
Taking advantage of this characteristic, we propose a measurement scheme for $\sim\mathrm{mT}$ static magnetic fields with a sensitivity of $10^{-7}~\mathrm{T/\sqrt{Hz}}/\sqrt{N}$ for $N$ nuclear spins. 
Therefore, the $10^6$ NV centers involved in the measurement can achieve a measurement sensitivity of $10^{-10}~\mathrm{T/\sqrt{Hz}}$, and the spatial resolution can reach the micron-scale.
Since the Hamiltonian of the electron in the rotating coordinate system is similar to that of the nucleus, our theory can be extended to the case of electrons. 
For the rotation frequency $\omega_\gamma\sim \mathrm{GHz}$, a single electron spin can achieve a sensitivity of $10~\mathrm{pT/\sqrt{Hz}}$. 
Besides, it would be interesting to investigate quantum metrology based on rotation induced non-adiabatic geometric phase in future.

\section{acknowledgments}
We thank Qing Ai for helpful discussions. This work is supported by the National Natural Science Foundation of China (Grant No. 12441502), and the Beijing Institute of Technology Research Fund Program under Grant No. 2024CX01015.

\bibliography{apssamp}
\end{document}